**Magnetically Controlling the Explosion of Dirac Fermions during the Oxidation of Graphene**

Reginald B. Little*, Glenn J. Gex, Muhammad Ansari and James W. Mitchell

Howard University

Crest Nanoscale Analytical Sciences Research and Education Center

Department of Chemical Engineering

2300 Sixth Ave NW, Washington, DC 20059

**Abstract**

The different physical properties of multilayered graphene or graphite relative to single layer graphene result from the Dirac spins symmetry in graphene and the Pauli spin symmetry in graphite. The Dirac symmetry in multilayers of graphene (graphite) is hindered by interlayer interactions. Different magnetizations, electronics and chemistry of graphite and graphene follow from absence of interlayer interactions in graphene. The distinct kinetics and dynamics of graphite and graphene by oxidation by the Hummer's method in weak external magnetic field are observed in this work. Graphite manifest enhanced non-explosive oxidation of Pauli spins in weak magnetic field with background paramagnetic oxygen slowing the magnetic acceleration. Graphene and graphite oxide manifest explosive oxidation and magnetically decelerated explosive oxidation of Dirac spins in weak magnetic field for temperatures below 20 $^{o}$Celsius. The explosive oxidation of graphene and its deceleration in weak external magnetic field are interpreted resulting from the giant nonlocality and spin Hall Effect in the chemically reacting graphene. This is the first identification, analysis and interpretation of the chemistry of the Dirac spins and the magnetochemistry of relativistic electrons.

* Corresponding Author; Reginald B. Little; redge_little@yahoo.com

**Introduction**

The objective of this work is to demonstrate the different chemistry of oxidation of graphene and graphite and to demonstrate new chemical kinetics and dynamics of graphene and graphite oxidations in weak magnetic fields on the basis of the different spin symmetries of Dirac and Pauli spins, respectively. Graphene is a single layer crystal of $sp^2$ bonded carbon atoms. Graphite is a composite structure of multilayers of strongly interacting graphene. The stacking of graphene layers in graphite leads to $p_z$ interlayer interactions. Diamagnetism and ferromagnetism of $p_z$ electrons in motions in each layer exert interlayer forces between layers. Such internal interlayer interactions by magnetism in graphite excite Landau levels internally in adjacent layers for shifting the electron spin symmetry from Dirac symmetry of graphene layers to Pauli symmetry in graphite. Such monolayer versus multilayer structural differences and consequent distinct intralayer and interlayer interactions (respecitively) lead to differing properties in graphite and graphene, respectively. Such interlayer magnetic interactions compare to the magnetic plasma and catalyst organizing graphene and CNT [1] by: dressing fermion masses [2]; achieving relativistic symmetry [3]; creating space-time [4]; pseudo spin and pseudomagnetism [5]; combining/separating electromagnetic and weak forces [6]; altering time reversal symmetry [7]; and mixing hybridization and bending orbitals [8]. Such interlayer interactions contribute to magnetic induced bandgap opening in biliayer graphene [9]. With more and more layers, the consequent interlayer interactions cause property differences such as the different transport properties in graphite relative to the electronic transport properties in graphene [9,10]. Multilayer graphene or graphite possesses spins of the Pauli type [11] due to the interactions between the layers, whereas single layer graphene possesses spins of the Dirac type [9,12]. The internal magnetism and hybridization of carbon atoms in graphene contribute to bonding and structure of symmetry described more by Dirac's Equation than Schrödinger's Equation. The resulting relativistic like symmetry of electrons in graphene exhibits unique physical properties, and chemical properties with unique magnetic alterations of such properties in graphene relative to graphite and its Schrödinger description. The multilayered graphite therefore manifests different properties in external magnetic field relative to the properties of single layer graphene in external magnetic fields [9,14].

The distinct physical properties of graphite and graphene have been more observed, reasoned and explained, but the differing chemical properties of these carbon structures have not been as fully explored. But a chemical reaction is an electron transfer with current on the atomic and molecular scales. So just as electronic differences exist between graphite and graphene, chemical differences should also exist. The distinct chemical bonding in graphene and graphite is manifested by varied Raman spectra of the graphite and graphene [15]. The differing chemical kinetics of thermal combustion of monolayer and bilayer graphenes in oxygen have been observed [16]. The different chemistry of graphite and graphene is here shown due to different spin symmetries of Dirac symmetry in graphene and Pauli symmetry in graphite. Graphite oxide manifests different chemistry relative to graphene oxide. The first experimental report of this was presented in a US Patent from 2009 [17]. Graphene nanoribbons even possess a different chemistry relative to macro graphene due the edges in nanographenes. These chemical differences between graphene, graphite and their oxides are expected to present different magnetic behaviors during chemistry and different chemical responses in external magnetic fields.

In considering the various physical differences, one of the many new interesting properties of graphite and graphene and even other carbon structures is ferromagnetism. Ferromagnetism in graphite differs from ferromagnetism in graphene. Ferromagnetism in graphene has been associated with the edges, defects, changes in hybridization and deformation [18,19]. Ferromagnetism in graphene involves Dirac Spins and is due to long range interactions and correlations of spins and pseudospins. Both spin and pseudospin motions contribute to ferromagnetism in graphene. The ferromagnetism in graphene is more coupled to electric force by the spin Hall Effect [20,21]. Therefore it is expected that Coulombic chemical potentials of reactants less effectively attack graphene relative to graphite. Pseudospins in graphene are collectively delocalized [5,21,22]. Currents magnetize graphene but do not magnetize graphite. Because chemical reactions are electric currents, it is expected that chemical reactions magnetize graphene over large space-time but not graphite [1,20]. The rapid electron motion in graphene contributes to higher magnetization of graphene relative to other substances. On the other hand, ferromagentism in graphite has been associated with defects [23], different hybridizations [24] and orbital motions about defects [25]. The Pauli spin, interlayer exchange and correlation may contribute to ferromagnetism in graphite. The magnetism in graphite is less coupled to electric force so the Coulombic chemical potentials of reactants more effectively attack graphite. The spins in graphite are more localized. On the basis of these distinct magnetisms in graphite and graphene, it is reasoned that they would manifest different physical and chemical properties and different effects on these properties in external magnetic field. In particular, it is expected that the influence of oxidation on the magnetization of graphene and on the influence of magnetic field on the oxidation of graphene will differ from that of graphite.

The oxidation of graphite by redox reactions with potassium chlorate and nitric acid was reported by Brodie in 1859 [26]. Brodie noted that upon heating the graphite oxide to 200 $^{\circ}$C , it released carbonic acid and carbonic oxide. In 1898, Staudenmaier increased the oxidation of graphite by Brodie's method by adding the chlorate in multiple steps (rather than in a single step) and using concentrated sulfuric acid [27]. By using multiple oxidations, Staudenmaier had unknowingly demonstrated oxidative differences in graphites of different thicknesses. Here we note the more Coulombic driving force of the Brodie and Staudenmaier methods relative to the subsequent Hummers' method. In 1958, Hummers and Offeman discovered another, safer oxidation method (known as Hummers' Method) for graphite oxidation using $NaNO_3$, $KMnO_4$ and $H_2SO_4$ [28]. Hummers' method substituted $KMnO_4$ for $KClO_3$. Here we note the less Coulombic and more magnetic nature of the Hummers' method in using $MnO_4^-$ relative to the Brodie and Staudenmaier methods of using $ClO_3^-$. In this work, the greater intrinsic magnetism of the Hummers' method relative to the Brodie and Staudenmaier methods is explored. These are the main three methods of chemically oxidizing graphite with achieving similar levels of oxidation but variations in the natures of the oxide products [29]. Recently, Tour's substituted $H_3PO_4$ for $HNO_3$ in the Hummers' method and reported improved efficiency of oxidation with less generation of toxic gases and easier temperature control [30].

The mechanisms of all four graphite oxidations are unknown. Here we report that the mechanism of oxidation should involve: the separation of the layers in graphite; the collective breaking of multiple π bonds and collective breaking of multiple σ bonds; the collective electron transfer to multiple oxidants; the chemical etching of macro-sized graphite and graphene to nano-size structures; the creation of radicals; the edge induced distortion, strain and stressing of C-C bonds; the rehybridization and bond rearrangements of the initial $sp^2$ carbon atoms of the layers in graphite and some formations of CO, $CO_2$, $CH_4$, and $C_2H_4$, $C_3H_4$, as well as other epoxide, hydroxyl, carbonyl and carbonic groups in the graphite and graphene. The chemistry during such mechanism for both graphite and graphene requires concerted reactions (or chemical currents) over different space-times with organization over larger space-time for graphene. The unusual electronic transport and thermal properties, giant nonlocality, quantum electrodynamics and Dirac symmetry are here noted to facilitate such organization of chemical dynamics over large space and time. The chemistry during such C-C bond breakages and C-O and C-H bond formations introduces H and O into the graphite and graphene lattices with triggering of ferromagnetism over larger space-time in the graphene in the various intermediates and products for alterations of the chemical kinetics and dynamics. Long range magnetization of graphene is thereby reasoned to organize these new chemical dynamics. The formation of magnetic moments in chemical intermediates during such oxidation dynamics is therefore within reason with stronger and longer range magnetization in graphene.

For all three methods of Brodie, Staudenmaier and Hummers, the reactants produce magnetic intermediates: Brodie's method ($NO_3^-$, NO, $NO_2$, and $N_2O$); ($ClO_3^-$, $ClO^-$, and $O_2$) and Hummers' method ($MnO_4^-$, $Mn_2O_7^-$, $MnO_3$, $MnO_2$, and MnO). The consideration of various oxidants used in the history of oxidizing graphite leads to increased magnetic nature of the redox reactants, intermediates and reactions from Brodie to Hummers. In particular, the shift from less magnetic chlorate oxidation and its explosive nature to more magnetic permanganate and its less explosive nature provides hints that intrinsic magnetism may be involved and magnetism may be a means of controlling oxidation of carbon via thermal, chemical and catalytic methods for safer and less explosive Hummers' method. In this work for the first time, we show that magnetization slows the explosive oxidation of graphene. Such magnetic mechanism of oxidation and its organization and synchronization of multiple electron transfer reactions may be extended to explain the explosive tendency of trinitrotoluene and nitroglycerine. In this investigation, we explore different magnetic environments on the Hummers' method. The oxidation of graphite and graphene by Hummers' method was monitored by observing $CO_2$ production under explosive and nonexplosive conditions, different temperatures and different permeability of background gas. Thereby the differing ferrochemical properties of graphite and graphene are determined [17] and the different effects of weak external magnetic field on these chemical properties are demonstrated [17].

**Experimental Section**

The novel chemical kinetics in graphene relative to graphite was observed by using the Hummers' method to oxidize graphite and graphene in the absence and in the presence of an applied weak external magnetic field (0.5 Tesla). The Hummers' method [28] involved weighing 0.4 g of carbon sample with 5.2 ml of concentrated sulfuric acid ($H_2SO_4$). 0.2g of $NaNO_3$ was then added followed by the addition of 1.2 g of potassium permanganate ($KMnO_4$). The reacting mixture was mixed well in a test tube and sealed with a pyrex connector tube. The sealed test tube was fixed between the poles of an electromagnet. The inlet to the test tube was connected to flow meter to control flow rates of carrier and background gases of air, argon, oxygen and nitrogen. The outlet of the test tube was connected to a residual gas analyzer. The mass spectra of the outlet gases were recorded in time as the oxidation-reduction reactions proceed to observe the mass intensity of various gaseous reactants and products of the Hummers' method. The kinetics of the Hummers' oxidation-reduction reaction were monitored by measuring the time evolution of a product ($CO_2$) produced by the oxidation-reduction reaction. The background carrier gas was changed from air to Ar-$O_2$ (50%-50%) to pure $O_2$ to alter the permeability of the gas background and analyze possible aerobic nature of the reaction. The carbon source was changed from graphite powder to graphene oxide to expanded intercalated graphite (graphene) to analyze the effects of multilayers versus single layer of carbon (reductant) reactant on the oxidation kinetics of the Hummers' Method. Conditions of the oxidation were varied so as to observe the products generated during nonexplosive oxidation and explosive oxidation of graphite in nonmagnetic and magnetic environments.

**Results and Discussion**

The observed kinetics by the amount of $CO_2$ production in time are presented in the various mass histories of oxidations of graphite and graphene in Figures I-V. In Figure I, the oxidation of graphite occurred in Ar-$O_2$ (50%-50%) atmosphere with an initial temperature of 25 $^o$C. In Figure I, it is observed that the application of an external magnetic field of 0.5 Tesla slightly accelerated the oxidation of graphite relative to the rate in zero applied external magnetic field as observed by the comparative rates of $CO_2$ production. The data in Figure 1 are consistent with the non-intercalated graphite being more readily magnetized by interlayer interaction under an applied external magnetic field. The external magnetic field interacts with Pauli spins in such multilayers (graphite) for magnetization of graphite. Multilayer graphite based on its electronic structure and properties has spins of the Pauli type and the interactions of these spins between layers leads to greater magnetization due to interlayer interactions and consequent gap opening in the multilayer relative to the single layer. The interlayer magnetization and gap opening torque electrons by the Little Effect for explaining the indirect bandgap nature of graphite. Such magnetization of graphite in the external magnetic field assists orienting the reducing graphite with the oxidants. As the reaction proceeds, the magnetic moments increase in the graphite. The magnetization of the graphite intermediates and oxidant intermediates may slow the reaction by the radical pair effect, but acceleration of the oxidation is observed instead. Just as the internal magnetization and interlayer interactions cause indirect transitions, the internal magnetization and interlayer interactions cause alteration of orbital symmetry of transferring electrons to accelerate the redox reactions. Such internal magnetism and interlayer interactions is here noted to alter the electronic orbital symmetry by the Little Effect for affecting a transfer of electrons from $sp^2$ orbitals of graphene into sp orbitals of $CO_2$ for allowing a nonpreservation of orbital symmetry during the chemical reaction contrary to the Woodward-Hoffman Rule [31]. A further acceleration of reaction in graphite in external magnetic field may be due to the Pauli spins in the graphite under magnetic orientation with the oxidants becoming more subject to the electrochemical potential of the oxidants relative to the weaker electrochemical action of the oxidants upon the Dirac spins in graphene (to be discussed later). Therefore the magnetic field couples in a different way to the multilayers of graphene in graphite and organizes, orients, rehybridizes and synchronizes the redox reactions with the oxidants for faster oxidation-reduction reactions relative to the lesser magnetic orienting of reactants and lesser inter-electrochemical action with a single graphene layer (to be considered below), wherein the spins behave like Dirac spins. Graphite has strong internal fields between layers but as graphite intercalates to graphene the field energy goes to kinetic energy of the electrons or relativistic motion of electrons in the expanded graphene layers. Graphite responds to oxidants Coulombic field by forming electric dipole but later graphene will be shown to form internal magnetic field to oppose the Coulombic fields of oxidants.. The Pauli spins in graphite also more strongly couple with the Coulombic fields of the oxidants for faster oxidation in the orienting external magnetic environment relative to the less orienting nonmagnetic environment. Such stronger coupling in magnetic field of graphite and the oxidants causes the observed faster oxidation kinetics of the graphite in external magnetic field relative to graphene under the applied external magnetic field.

The observed kinetics by the amount of $CO_2$ production in time are presented for the non-explosive oxidation of graphite in pure oxygen background in Figure II. The oxidation of graphite by the Hummers' method was done with an initial temperature of 25 $^oC$. The oxidation in pure oxygen occurred nonexplosively as observed by the slower production rate of $CO_2$ in Figure II. The graphite oxidation in pure oxygen was slightly faster for the nonmagnetic relative to the magnetic as observed in Figure II. Changing from dilute $O_2$ in Figure I to pure $O_2$ background in Figure II caused a dramatic slowing of the oxidation in magnetic field due to the paramagnetic nature of $O_2$ and the increased $O_2$ concentration increasing the permeability and spin polarization of the gaseous background there by diminishing the magnetic field on the reactants and their orientations and spin dynamics of their chemistry. But comparing the influence of pure $O_2$ on the graphite oxidation in nonmagnetic field, it is observed that the pure $O_2$ slightly increased the oxidation rate, which may be associated with $O_2$ magnetizing the graphite [32] for the resulting internal magnetization increasing the graphite oxidation. $O_2$ interacts with strong internal fields in graphite to magnetize graphite. It has been has also been shown that $p^+$ interact with the internal fields in graphite to magnetize it [33]. The fine structure in the mass spectrum reflects the oxygen altering the spin exchange and spin torque during the electron transfer. The $O_2$ is transferring away pseudospin magnetism. In external magnetic field, the $O_2$ is organized to transfer more pseudospin magnetism away during the oxidation. But increasing $O_2$ background causes higher permeability which diminishes the external magnetic field acceleration of the graphite oxidation. So diluting the external magnetic field by using pure $O_2$ caused the diminished magnetic orientation on the graphite and the oxidants but moreover the surrounding spin polarized $O_2$ slowed spin exchange and spin torque of transferred electrons from the reacting graphite and to oxidants for slowing of the oxidation-reduction relative to non-applied magnetic oxidation-reduction. Magnetic field acts on graphite, oxidants and $O_2$ in dilute $O_2$, but in concentrated $O_2$ the permeability is higher, so reactants are shielded from external magnetic field with a loss of orienting effect of the external magnetic field on reactants. So the magnetic field is not able to orient the reactants as well as in Figure I (dilute $O_2$) and the reaction rate in the external magnetic field is slower in pure $O_2$ (Figure II) relative to the diluted $O_2$ background (Figure I). The spin polarized $O_2$ background may alter the interlayer torque of electrons azimuthal rehybridization (the Little

Effect) as they are transferred from graphite to oxidants. In addition to this change in permeability in pure $O_2$, the $O_2$ adsorption on graphite and the various intermediates enhances the ferromagnetism of the graphite and graphite oxide for causing magnetic slowing of the reaction of graphite and graphene with the oxidants. The magnetized $O_2$ in the external magnetic field is less able to exchange magnetization of the oxidation reaction with a further slowing relative to nonmagnetic oxidation environments for the observed faster oxidation in nonmagnetic environment.

The observed kinetics by the amount of $CO_2$ production in time is presented for the explosive oxidation of graphite oxide in Figure III. The explosive oxidation of the graphite oxide was done in Ar-$O_2$ (50%-50%) atmosphere with an initial temperature of 25 $^oC$. The graphite oxide oxidation (Figure III) occurs more explosively than graphite oxidation (Figure I). The oxidations of graphite oxide in magnetic and nonmagnetic external fields are faster than oxidations of graphite in both magnetic and nonmagnetic fields in Figures III and I, respectively. The faster oxidation of graphite oxide is due to the isolated monolayers in graphite oxide and consequent Dirac symmetry of spins in the isolated monolayers of the graphite oxide relative to Pauli symmetry of spins in graphite. In graphene and graphite oxide, the huge energy of the internal fields in graphite of interlayer interactions is converted to high speed of the electrons in the monolayer (Dirac spins). For the graphite oxide, the oxidation in applied magnetic field is initially faster than oxidation without applied magnetic field in Figure III. But later during the oxidation, the rate in applied magnetic field slows relative to the rate of oxidation in nonapplied magnetic field in Figure III. The initially faster oxidation in magnetic field may be an orienting effect and an effect of loss of oxide groups and graphene formation.

The oxidation of the isolated single layers in graphite oxide is explosive and faster than the non-explosive oxidation of graphite in Figure I due to the isolated layers of the graphite oxide and the resulting Dirac chemistry in comparison to the multilayers for graphite and graphite's Pauli chemistry. Graphite oxide may be more ferromagnetic than the graphite powder due to effect of $O_2$ adsorption [32]. The graphite oxide may be more magnetic than the intercalated graphite (graphene) for slower oxidation of graphite oxide relative to isolated graphene due to the greater initial internal ferromagnetism in graphite oxide and consequent magnetic slowing of its oxidation. The Dirac nature of spins in graphene oxide single layers and their collective dynamics contributes to giant nonlocality and giant spin Hall Effects with the consequent huge internal magnetic field and electron interactions in the monolayer for collective magnetically synchronized electron transfer to oxidants for chemical explosions during oxidation by Dirac chemistry. Such collective multiple electron transfers from graphene and graphite oxide are consistent with the observed electric sparking during the chemistry of the Hummers' method. An applied magnetic field can better control and interact with the collectively long range interacting Dirac spins in the isolated layers in the graphite oxide and graphene to thereby magnetically slow the collective electron transfers from graphite oxide and graphene to oxidants and to thereby magnetically slow the explosive collective electron transfer chemistry in the external magnetic field. In graphene oxide and graphene external magnetic field is controlling Dirac $\pi$ spins as they spin polarize and better ignore and resist Coulomb field of the oxidants in the external magnetic field. External magnetic fields create a large dipole perpendicular to the graphene in $p_z$ for polarizing and creating large magnetic field internally in graphene and graphite oxide. As the Coulombic field of the oxidants increases the graphene and graphite oxide respond by generating greater magnetization to oppose the chemical potential of the oxidants. In graphite oxide and graphene, the external magnetic field also slows the multiple electron transfer by spin polarizing the wave function in graphene with the wave functions in the oxidants. Such magnetic diminish of the Coulombic attraction of Pauli spins in graphite does not occur (Figure I) as it occurs in graphite oxide and graphene. Such slowing of Dirac spins of graphene and weakening of the Coulombic pull on Dirac electrons in graphene by the Coulombic fields of oxidants slow oxidation kinetics of graphene relative to graphite in the external magnetic field and also explains less multiple electron transfer to oxidants for less $CO_2$ formation. Isolated graphite oxide and graphene are initially magnetically oriented faster by the external field than the orienting of graphite, but then the multiple electron transfer to oxidants from the oriented graphene oxide and graphene is decelerated (relative to graphite) in external magnetic field due to Dirac symmetry of $\pi$ electrons and their excitation and magnetization in the weak external magnetic field and their greater blindness to the Coulomb potential from the oxidants.

The observed kinetics by the amount of $CO_2$ production in time are presented for the explosive oxidation of expanded graphite (graphene) in Figure IV. The explosive oxidation of the expanded graphite (graphene) was done in $O_2$ and Ar at initial temperature of 18 $^o$C. The acid pretreatment expanded the graphite for more isolated graphene layers and caused more reproducible explosive oxidation of isolated graphene in $O_2$ within the diluting atmospheric amounts of $N_2$ and $O_2$. It has been demonstrated that acid pretreatment intercalates and forms bilayer and expands graphite for graphene and expanded graphite formations [34,35]. Recently after this current work, other researchers have reported computational data that are consistent with our experimental data reported here wherein $H_2SO_4$ breaks bilayer graphene into monolayer graphene with the resulting monolayer graphene having similar properties as free-standing graphene [35]. These subsequent computational results of $H_2SO_4$ exfoliating bilayer graphene into monolayer graphene are consistent with the prior experimental observations of pretreated multi-layer graphene forming monolayer graphene and the Dirac explosions reported in this manuscript. Consistent with these computational findings and the experimental data for sulfuric acid exfoliations of graphite down to monolayer graphene reported here, Esquinazi et al [36] report more experimental data of sulfuric acid inducing ferromagnetism in the surface layers of graphite due to the protons slightly altering the pi bonds in the first few layers in graphite. It is important to note the greater reproducibility of explosive oxidation at initial temperature of 18 $^o$C relative to the poorer reproducibility of explosive oxidation at initial temperature > 25 $^o$C. This greater explosion at lower temperature correlates with the observed anomalous quantum spin Hall effect in graphene for temperatures less than 22 $^o$C [21,22]. It is important to consider the consistency between graphite oxide and graphene as the explosion in graphite oxide occurred for T > 25 $^o$C but the explosion in graphene required T < 22 C. The greater ferromagnetism initially in graphite oxide explains graphite oxide's ability to explode at higher temperatures relative to explosion of graphene at lower temperature of 18$^o$C. This is the first observation of the anomalous quantum spin Hall Effect and giant nonlocality of spin polarization and magnetization on the chemical reaction of graphene in weak magnetic fields. For such explosive oxidation, the graphene explosion with no applied magnetic field exhibited faster oxidation kinetics than the graphene explosion in applied magnetic field in Figure IV. The expanded graphite (graphene) behaved similar to the graphite oxide in its oxidation kinetics, but the oxidation rate is faster for the expanded graphite (graphene) relative to the graphite oxide.

The expanded graphite (graphene) caused explosive chemistry during oxidation due to the Dirac nature of the spins in the graphene. Unlike the unexpanded graphite which exhibits Pauli spin behavior the more isolated layers in the expanded graphite exhibit more Dirac nature and more explosive chemistry: Dirac chemistry. The Dirac chemistry and explosiveness in the monolayer graphene is countered by interlayer interactions and magnetism with the emergence of Pauli type spins and nonexplosive chemistry in graphite as explosive mutually layers interact to counter their strong interactions and weaken explosion relative to the monolayers. This greater explosion of monolayers and its interpreted Dirac electronic origin rather than a transport cause is explicitly experimentally demonstrated by further comparing the graphene and graphite oxidations to the oxidation kinetics of graphite oxide. If the greater explosiveness of graphene relative to graphite was due to simply better transport and better accessibility of the graphene monolayer to oxidizing reactants then graphene oxide should have faster kinetics of oxidation than pure monolayer graphene as the graphite oxide is just as isolated and single layered and already partially oxidized relative to the expanded graphite or graphene. But in comparing Figures IV and V, the pure graphene and the graphite oxide oxidations reveal faster oxidation kinetics of the pure graphene. Therefore the observed faster kinetic explosion in the graphene is not due to transport effects but is a result of intrinsic electronic differences of Dirac symmetry in graphene and Pauli symmetry in graphite. The fewer explosions in the magnetic field follow from the field controlling Dirac fermions collectively in the reductant (graphene) for the magnetization of the Dirac spins in the weak magnetic field and Coulombic potential of the oxidants for less rehybridization of the delocalized Dirac spins of graphene to localized Pauli spins (sp) of $CO_2$ for less consequent multi-centered transfer to multi oxidants for explosion. Just as the external magnetic field interacts with the Dirac spins in the monolayer graphene, surrounding paramagnetic $O_2$ also can interact to slow the Dirac spins in the monolayer graphene to alter is properties. Subsequent to this work other researchers have observed the enhanced superlinearity of the dispersion in graphene in a vacuum where there is no $O_2$ [37]. Also the ability of graphene to ignore pure coulombic potential of reactants but the ability of the graphene to electrochemically respond to magnetic oxidant ferrycyanide in an electric field has been demonstrated [38]. The oxidation of the Dirac spins in the nonapplied magnetic field of the expanded graphite is faster due to less orientation less hindered internal magnetic rehybridization and the greater response of the Dirac spins in expanded graphite to the internal magnetism and Coulombic interactions of the $Mn_xO_y$ and $N_xO_y$ oxidants when no external magnetic field is applied. The oxidants by the Hummer method present a Coulombic and magnetic interaction with the graphene Dirac spins, the magnetic interactions of the $Mn_xO_y$ and $N_xO_y$ species pull the Dirac spins in the graphene. It is interesting to compare such magnetic and magnetically correlated oxidants with paramagnetic oxidants as used by Flynn and Brus [16]. The weaker paramagnetic oxidant $O_2$ requires higher temperature (250 $^o$C) for oxidation of monolayer graphene with different onset reaction temperatures for bilayer and trilayer graphene and the $O_2$ lacked exchange and ferromagnetism for explosive oxidation as can occur for $MnO_4^-$ and $NO_3^-$ in the Hummer method. The $Mn_xO_y$ and $N_xO_y$ complexes organize magnetically the concerted transfer of many electrons from graphene where as the $O_2$ paramagnetic gas is not able to organize such explosive transfer. When a magnetic field is applied to the Dirac spins in the isolated graphene layers of the expanded graphite, the Dirac spins in graphene are more oriented with the Coulombic fields of the oxidants; more strongly influenced by the external magnetic field due to their Dirac nature; more internally magnetized by the increase Coulombic chemical potential of the oxidants (by giant spin Hall Effect) for magnetic opposing electron transfer; are less prone to rehybridization from Dirac to Pauli spin due to the large nonlocality; and are less coupled to the Coulombic and electrochemical fields of the oxidants and the internal magnetic moments of the oxidants for slower Coulombic driving force for multiple electron transfer and oxidation from graphene to $MnO_4^-$ and $NO_3^-$ to form $CO_2$ and graphene oxide. Dirac spins are more influenced by magnetic field of external magnet and internal magnetism of the oxidants, but the Dirac spins are not as influenced by Coulomb field of oxidants. As Coulomb field is applied to the graphene, it becomes more magnetized. As the Coulombic field of oxidants increases the magnetization of the graphene increases to slow and oppose the electron transfer to the nearby oxidants. The graphene in the weak external magnetic field and strong Coulomb field of oxidants manifests nonlocality of Dirac spins, separation and magnetization of pseudospins and giant spin quantum Hall Effects, which magnetically slow the multiple electron transfer of such magnetically polarized spins to the strong oxidants for magnetically slowing the Hummers' method. The external magnetic field slows the intrinsic internal pseudospin torque of graphene obitals from $sp^2$ delocalized symmetry into sp hybridized localized carbon orbitals of the $CO_2$ product. External magnetic field also slows the collective electron transfer from graphene to oxidants by magnetically polarizing the Landua levels on graphene with the orbitals of the oxidants ($Mn_xO_y$ and $N_xO_y$). So the external magnetic field more strongly alters the Dirac spins in the expanded graphite (graphene) thereby making the Dirac spins less accessible to the oxidants and slowing the oxidation of the expanded graphite in external magnetic field relative to the oxidation without an applied magnetic field. The external magnetic field also magnetically orients the magnetic oxidants and intermediates of the oxidants for stronger magnetization of Dirac spins in the graphene reductant and even greater resistance to chemical potential due to the oxidants. Furthermore, in nonmagnetic environment there was less spin alignment and less magnetic repulsions between intermediates relative to the greater internal magnetic repulsion and the magnetically hindered Dirac spins formed during oxidation in external magnetic field.

The observed kinetics by the amount of $CO_2$ production in time is presented for the non-explosive oxidation of expanded graphite in Figure V. The non-explosive oxidation of expanded graphite was done in $Ar-O_2$ (50%-50%) at initial temperature greater than 20 $^o$C. The higher temperature diminished the quantum mechanical effects and collectivity during the oxidation for fewer explosions. The non-explosive oxidation of expanded graphite is faster in nonmagnetic field relative to magnetic oxidation in Figure V. It is interesting to compare the nonexplosive oxidation of unexpanded graphite (Figure I) and the non-explosive oxidation of expanded graphite (Figure V). The unexpanded graphite had faster oxidation in magnetic field, where as the expanded graphite had faster oxidation in non-applied magnetic field. The expanded graphite has less interlayer interaction and lower magnetization than the un-expanded graphite so the expanded graphite is initially less magnetized and less orienting occurs and more Dirac nature of spins exist, which both slow spin transfer to the oxidants for slower oxidation of the expanded graphite in applied magnetic field. Furthermore the giant spin Hall Effect and nonlocality in the expanded graphite (graphene) causes opposing magnetization in the graphene as the Coulombic potential increases for the oxidants during the oxidation. The faster oxidation without a magnetic field for the expanded exploding graphite and the expanded nonexploding graphite is due to the Dirac nature of the spins and the initial weaker magnetization of the expanded graphite so that the increasing magnetization and the magnetically excited Dirac spins diminish the Coulomb pull from the oxidants to oppose the multiple collective electron transfer from graphene to the oxidants. Magnetize Dirac spins slow oxidation and magnetize Pauli spins accelerate oxidation. The magnetic field entangles Dirac spins over space time. The magnetic field polarizes Pauli spins and orient orbitals of Pauli spins.

In this graphite oxidation reaction by Hummers' Method, there should be a magnetic effect based on the radical pair effect as the magnetic field can couple to the reactants and radical intermediates to polarize spins of radicals and intermediates for alterations in chemical kinetics and dynamics. But the radical pair effect has typically been observed nonadiabtically. But what field strength is necessary to influence such oxidation reactions of carbon by the radical pair effect? Surprisingly we observe the effects on the oxidation-reduction chemistry by Hummers' method in rather weak magnetic fields (0.5 Tesla) but such weak magnetic field effects on the chemistry of graphene should not be a great surprise as recently giant spin quantum Hall effects and giant nonlocality for current induced magnetization of graphene were observed at weak magnetic fields [21,22]. The giant nonlocality and anomalous quantum spin Hall Effect are here shown to alter the oxidation of graphene by external weak magnetic field. Such correlation of weak magnetic field effect on the graphene oxidation and the quantum spin Hall Effect are consistent with observed maximum temperature of about 22 $^{o}$C for both quantum oxidation reported here both and for the quantum spin Hall Effect [21,22].

It is further crucial to consider that the ferromagnetism of carbon alters the magnetic field effect on carbonaceous substances in ways beyond the radical pair effect [1,20]. Here we observe new effects emerging for the oxidation of ferrocarbon in comparison to oxidation of other paramagnetic substances and the old radical pair effect. Just as new effects exist in the chemistry of ferrometals relative to other transition metals, new effects should manifest due to the exchange/correlation of multiple Pauli spins in ferrocarbon that do not exist in other paramagnetic substances and the radical pair effect [1,20]. Moreover new effects should thereby manifest during the chemistry of graphene due to the exchange/correlation of multiple Dirac spins in graphene and its magnetization under weak external magnetic and strong Coulombic fields of oxidants. The external magnetic field can alter the electron transfer from a semimetal to polarize such topological insulator like graphene by the external chemical potential magnetizing the graphene. One new effect is the Little Effect [1,20], where in strongly correlated spins alter their orbitals via spin alterations to alter the bonding symmetry and dynamics during chemistry. The external magnetic field can also excite collective motion of Dirac spins in reactants to create greater inertness to Coulombic intermolecular forces and chemical potentials for retarding chemical reactions as observed here. External magnetic field also orients macromolecules and nanostructures for novel effects. Macromolecules and nanostructures can have multiple spin centers with ferromagnetism of those centers for novel synchronized chemistry that does not occur in smaller paramagnetic molecules [1,20]. Just as magnetic field couple small paramagnets to form larger paramagnets and to form ferromagnetic carbon, external magnetic field can also organize the decomposition and oxidation of large diamagnets and ferromagnetic reactants and intermediates. External and internal magnetic fields in and on reactants and intermediates can break or create new entanglement during chemical bond rearrangements [1,20]. Such magnetic influence on entanglement is an emerging phenomenon different from radical pair effect in that the magnetic field couples unentangled reactants and entangle the atoms, molecules and nanostructures to form new products, whereas the old radical pair effect involves already entangled atoms with breaking a chemical bond and the loss of entanglement slowing the rebonding [1,20]. The strong magnetic field can accumulate radicals for multi radical interactions to magnetically alter the entanglement to form new products [1,20]. Such influence of magnetic field on graphene is altered in graphite as the layers interact to suppress the Dirac electronics observed in isolated graphene and affect more Pauli type symmetry of electrons in graphite. By separating the layers we observe novel chemical kinetics in the graphene relative to graphite and even in weak magnetic fields. Subsequent to this work [17], other researchers have observed similar accelerated oxidation of graphite by ferricyanide in electric field [38].

**Conclusions**

The oxidation kinetics of carbon exhibits differences in graphene (single layer) and graphite (multilayers). The different oxidation kinetics of carbon atoms in graphene and graphite may be interpreted on the basis of electronic and magnetic structural differences between graphene and graphite for Dirac type spin (relativistic) symmetry and Pauli type (non-relativistic) spin symmetry. Such electronic and magnetic structural differences further account for explosive oxidation of graphene by Dirac chemistry and nonexplosive oxidation of graphite by classic Pauli chemistry. Both graphene and graphite oxidations couple to external magnetic field with the different spin symmetries causing distinct different coupling kinetics and ferrochemistry and magnetic control of the ferrochemistry. The different spin symmetries in graphene and graphite and distinct coupling to external magnetic field also manifest different dynamics of oxidation. This is the first observation, identification and interpretation of such unique ferrochemical dynamics in graphene as related to the giant nonlocality and spin Hall Effect in graphene.

**Acknowledgement**

This work is supported by the proceeds from an Endowed Chair (by the David and Lucille Packard Chair Materials Research Fund) in the Department of Chemical Engineering at Howard University.

**Figure Legend**

**Figure I – Mass History of $CO_2$ Production during Non-Explosive Oxidation of Graphite in $O_2$-Ar (50%-50%)Atmosphere**

**Figure II – Mass History of $CO_2$ Production during Non-Explosive Oxidation of Graphite in Pure $O_2$ Atmosphere**

**Figure III – Mass History of $CO_2$ Production during Explosive Oxidation of Graphite Oxide in $O_2$-Ar (50%-50%)Atmosphere**

**Figure IV – Mass History of $CO_2$ Production during Explosive Oxidation of Graphene in $O_2$-Ar (50%-50%)Atmosphere**

**Figure V – Mass History of $CO_2$ Production during Non-Explosive Oxidation of Graphene in $O_2$-Ar (50%-50%)Atmosphere**

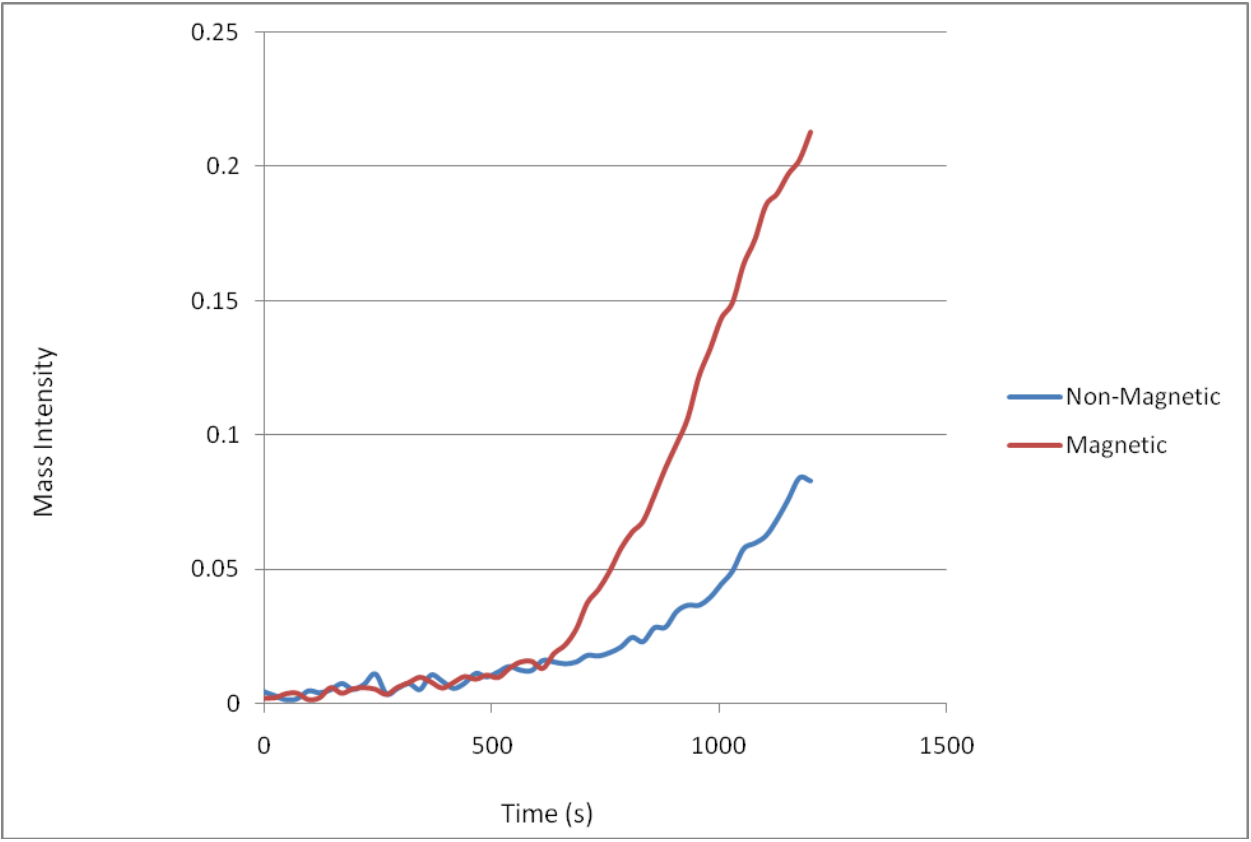

**Figure I**

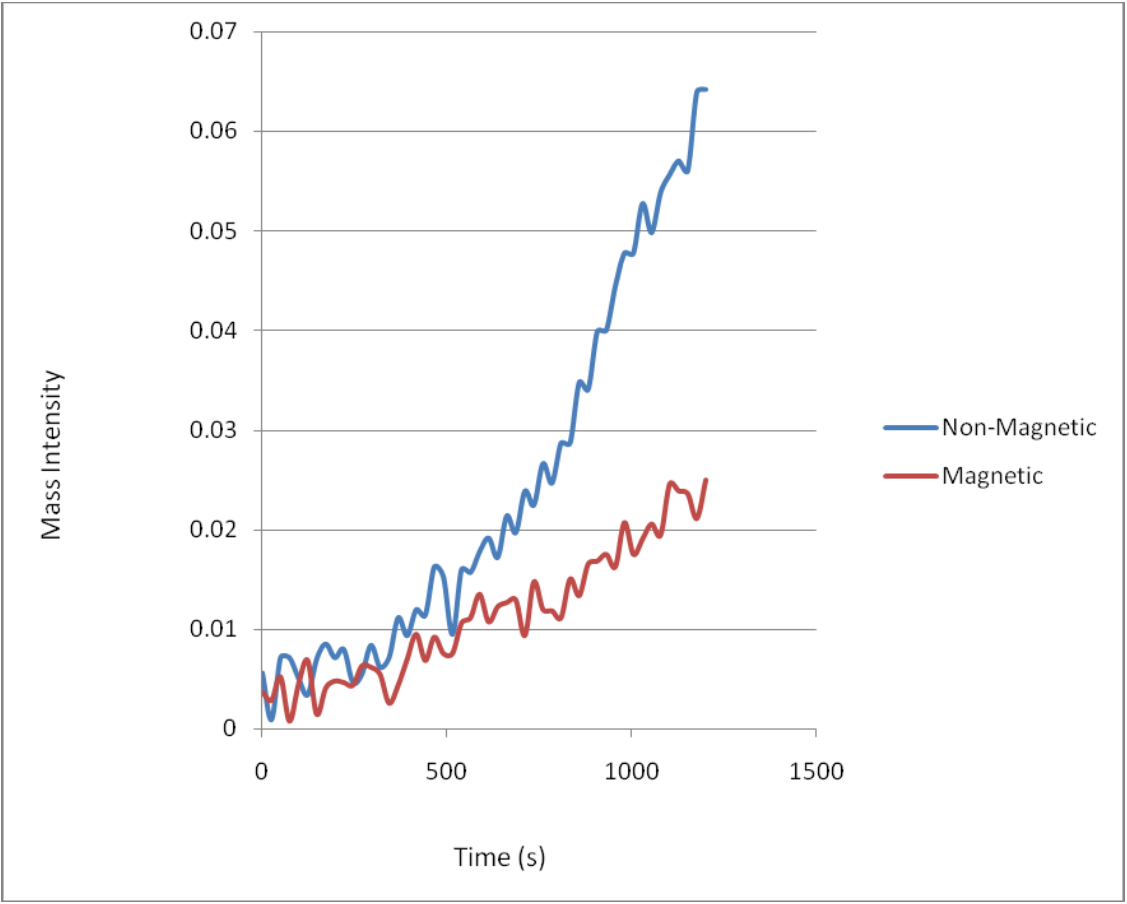

**Figure II**

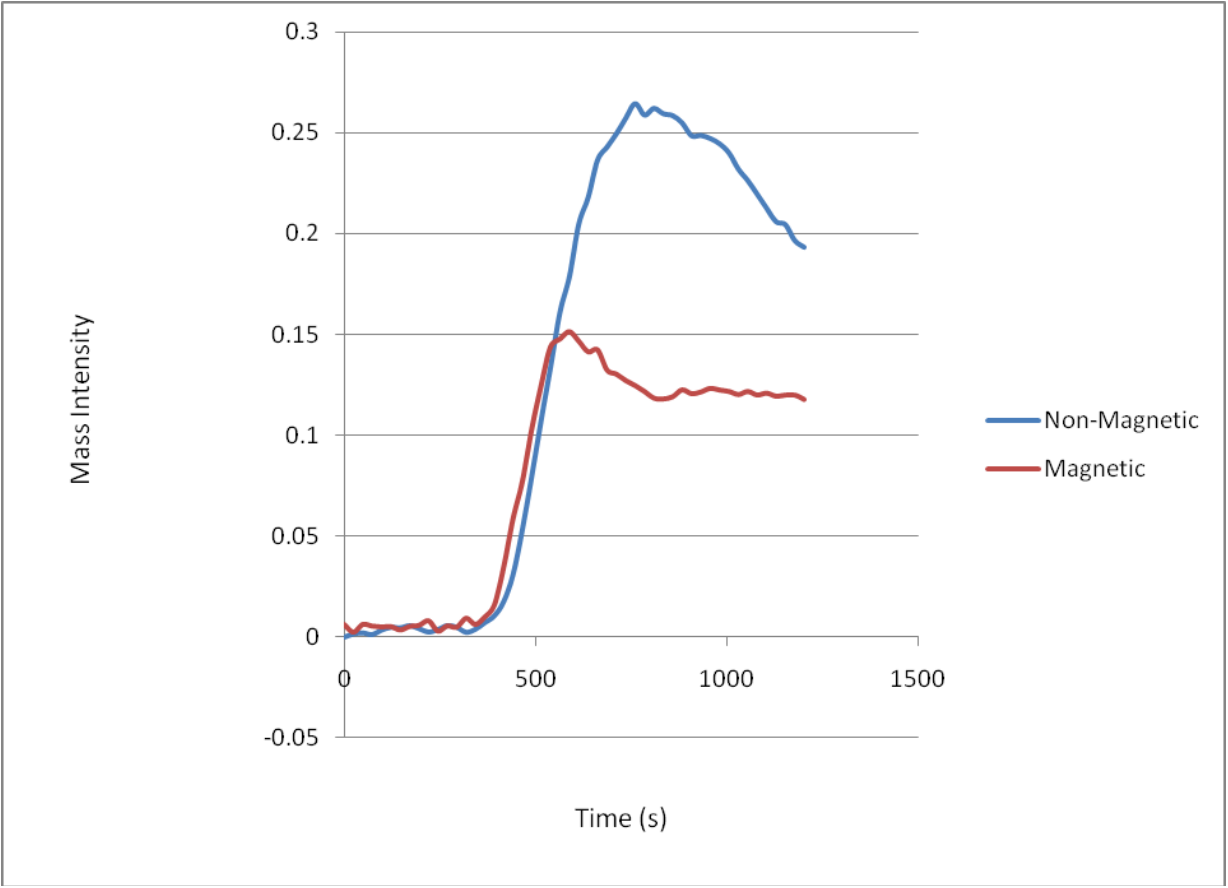

**Figure III**

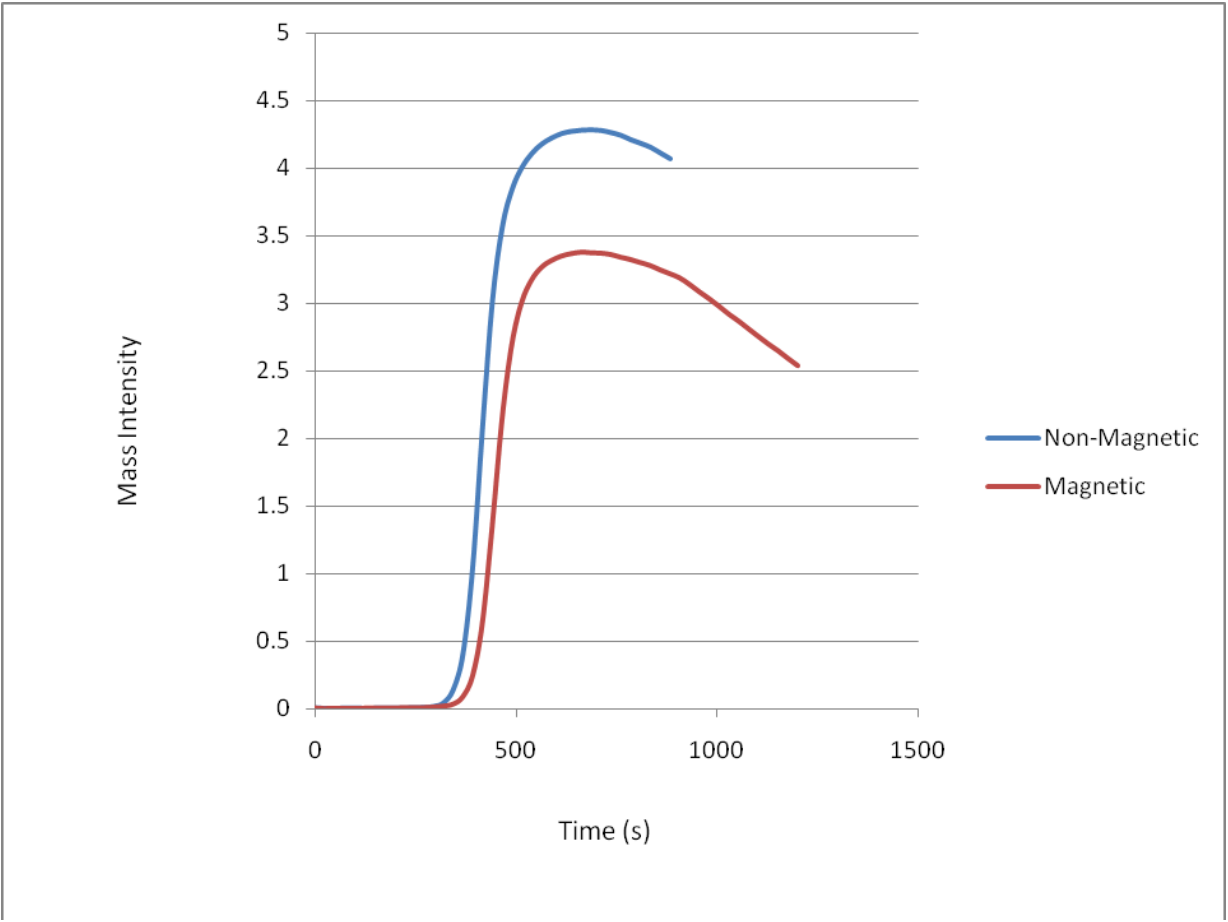

**Figure IV**

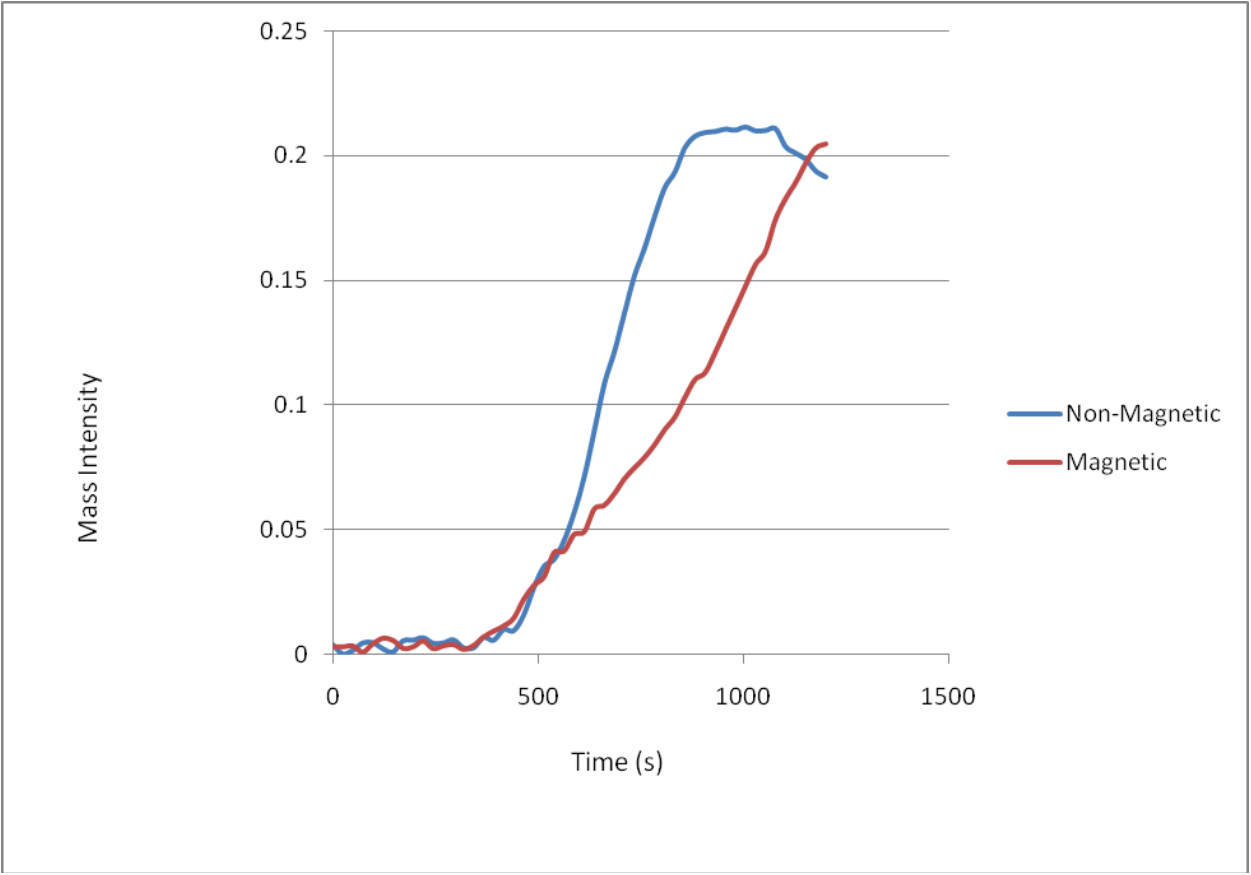

**Figure V**